\documentclass[aps,pra,preprint,groupedaddress]{revtex4-1} 

\usepackage{amsmath}
\usepackage{amssymb}
\usepackage{mathrsfs}
\usepackage{graphicx}
\usepackage{color}
\usepackage{enumitem}
\usepackage{caption}
\usepackage{subfigure}

\begin{document}


\title{Efficient quantum circuit for singular value thresholding}
\author{Bojia Duan}
\email[]{deja@nuaa.edu.cn}
\author{Jiabin Yuan}
\email[]{jbyuan@nuaa.edu.cn}
\author{Ying Liu}
\author{Dan Li}
\affiliation{College of Computer Science and Technology, Nanjing University of Aeronautics and Astronautics, No.29 Jiangjun Avenue, 211106 Nanjing, China.}

\date{\today}

\begin{abstract}

Singular value thresholding (SVT) operation is a fundamental core module of many mathematical models in computer vision and machine learning, particularly for many nuclear norm minimizing-based problems. A quantum SVT (QSVT) algorithm has been proposed for solving an image classification problem in \emph{Phy.Rev.A. 96, 032301,2017.} This algorithm runs in $O\left[\log_2\left(pq\right)\right]$, an exponential speed improvement over the classical algorithm which runs in $O\left[poly\left(pq\right)\right]$. In this paper, we design a scalable quantum circuit for QSVT. The quantum circuit is designed with $O\left[ {{{\log }_2}\left( {pq} /{\varepsilon} \right)} \right]$ qubits and $O\left[ {ploy\log_2 \left( {1/\varepsilon } \right)} \right]$ quantum gates in terms of error $O\left( {\varepsilon} \right)$. We also show that high probability and high fidelity output can be obtained in one iteration of the quantum circuit.
The quantum circuit for QSVT implies a tempting possibility for experimental realization on a quantum computer. Finally, we propose a small-scale quantum circuit for QSVT. We numerically simulate and demonstrate the performance of this circuit, verifying its capability to solve the intended SVT.

\end{abstract}


\maketitle


\section{Introduction}
\label{sec1:level1}

    Quantum computing has been shown to perform significantly better than classical computing at certain computational tasks, especially in the emerging interdisciplinary field of quantum machine learning \cite{NC10,SCM12,SSP14,Wittek14}. To show its superiority, remarkable quantum algorithms help a range of classical algorithms achieve a speedup increase \cite{Shor94,Grover97,DH96, BHM00,GLM08a,HHL09,LYD17,LMM16}. Shor's algorithm for factoring and Grover's algorithm for search are typical algorithms which can achieve exponential and quadratic speed increase, respectively \cite{Shor94,Grover97}. In 2009, Harrow, Hassidim, and Lloyd (HHL) proposed an algorithm for solving linear systems of equations \cite{HHL09}. This algorithm offers an exponential speed over its classical counterparts by calculating the expectation value of an operator associated with the solution of the linear equation under certain circumstances. Considering that a linear system is the centre of various areas in science and engineering, the HHL algorithm guarantees widespread applications \cite{LMR14,RSL16,RML14}. Inspired by the idea of the HHL algorithm, other fruitful quantum machine learning algorithms are proposed \cite{LZ16,SSP16,AAD15,QHL13,YGW16}. Moreover, experimentalists aim to implement the HHL algorithm on a quantum computer. Therefore, research on the numerical theoretical simulation and experimental realization of the algorithm is emerging recently \cite{CDF12,PCY13, CWS13, SKR14}.

    In many mathematical models in computer vision and machine learning, there is a fundamental core module known as singular value thresholding (SVT). In particular, the SVT method has been widely adopted to solve many nuclear norm minimizing (NNM)-based problems, such as matrix completion, matrix denoising and robust principle component analysis (RPCA). Therefore, the SVT method can be applied to many applications, such as image extraction, image colorization, image denoising, image inpainting and motion capture date recovery. However, NNM usually requires the iterative application of singular value decomposition (SVD) for SVT and the computational cost of SVD may be too expensive to handle data with high dimension, e.g. high-resolution images. In order to speed up the process of SVT, we proposed a quantum SVT (QSVT) algorithm that can execute the SVT operator exponentially faster than its classical counterparts. The QSVT algorithm is used as a subroutine to accelerate an image classifier SMM \cite{Duan17}.

    In this article,  we conduct further specific discussion regarding the algorithm and design a quantum circuit for QSVT which can be applied on a universal quantum computer.
    And the scaleable quantum circuit requires $O\left[ {{{\log }_2}\left( {pq} /{\varepsilon} \right)} \right]$ qubits and $O\left[ {ploy\log_2 \left( {1/\varepsilon } \right)} \right]$ quantum gates with error $O\left( {\varepsilon} \right)$.
    Ref. \cite{Duan17} shows that the QSVT algorithm is based on the HHL algorithm which consists of two core subroutines, namely, phase estimation and  controlled rotation. Phase estimation outputs the eigenvalues of input matrix ${\bf{A}}$ and decompose the input vector $\left|\psi_{A_0} \right\rangle$ in the eigenbasis of ${\bf{A}}$. Innovating and  implementing controlled rotation are the key of HHL-based algorithms. Herein, we add an important missing piece to the algorithm by developing the detailed circuit of the controlled rotation via the theoretical function in Ref. \cite{Duan17}.
    Specifically, we divide the controlled rotation into two unitary operations.
    The first unitary operation is ${\bf{U}}_{\sigma,\tau}$, which is used to compute the function of the eigenvalues of ${\bf{A}}$. By implementing the Newton iteration method, these function values can be stored in the quantum basis states. The second unitary operation is ${\bf{R}}_y$, which is used to extract the values in the quantum basis state to the corresponding amplitudes of the basis states. Implementing ${\bf{R}}_y$ directly affects the probability of obtaining the final result and the fidelity of actual and ideal final states. To improve probability and fidelity, we introduce a parameter $\alpha$ in ${\bf{R}}_y$, which can be computed ahead of quantum circuit implementation. Moreover, we present an example of a small-scale circuit for the algorithm and execute the numerical simulation of the example. The result shows the capability of the quantum computer to solve the intended SVT, and the performance of the algorithm is discussed.

    In detail, our work has two major contributions. First, we design a quantum circuit for QSVT algorithm, which provides a possibility for implementing the algorithm on a quantum computer. Second, by introducing the parameter $\alpha$, which can be computed ahead of implementing the quantum circuit, in the circuit design of controlled rotation, high probability and high fidelity can be obtained. Our work based on the QSVT algorithm may also inspire the circuit design of other HHL-based algorithms.

    The remainder of the paper is organized as follows: We give a brief overview of QSVT algorithm in Sec. \ref{sec2:level1}. Sec. \ref{sec3:level1} puts forward the quantum circuit for QSVT and analyzes the probability and fidelity. We propose an example in Sec. \ref{sec4:level1}, and present our conclusions in Sec. \ref{sec5:level1}.

\section{Review of QSVT}
\label{sec2:level1}
In this section, we briefly review the QSVT problem and the key procedure of QSVT algorithm. More detailed information can be found in Ref. \cite{Duan17}.

\subsection{QSVT problem}
\label{sec2.1:level2}

	SVT is an algorithm based on the SVD of a matrix. Suppose the input of the SVT is a low-rank matrix ${\bf{A}}_0 \in \mathbb R{^{p \times q}}$ with singular value decomposition ${{\bf{A}}_0} = \sum\nolimits_{k = 1}^r {{\sigma _k}{{\bf{u}}_k}{\bf{v}}_k^T}$, where $r \le \min \left( {p,q} \right)$ is the rank of ${\bf{A}}_0$, and ${\sigma _k} \left( {{\sigma _1} >  \cdots  > {\sigma _r} > 0} \right)$ are just the singular values of ${\bf{A}}_0$, with ${{\bf{u}}_k}$ and ${{\bf{v}}_k}$ being the left and right singular vectors.
     	SVT solves the problem ${\bf{S}} = \mathcal{D}_\tau \left( {{{\bf{A}}_0}} \right): = \sum\nolimits_{k = 1}^r {{{\left( {{\sigma _k} - \tau } \right)}_ + }{{\bf{u}}_k}{\bf{v}}_k^T} $, where ${\left( {{\sigma _k} - \tau } \right)_ + } = \max \left( {{\sigma _k} - \tau ,0} \right)$ and $\tau \in (0,\sigma_1)$ \cite{CCS08}.
     	The vectorization of the matrices ${{\bf{A}}_0}$ and ${\bf{S}}$ are ${{\rm{vec}}({\bf{A}}_0^T)} = \sum\nolimits_{k = 1}^r {{\sigma _k}{{\bf{u}}_k} \otimes {{\bf{v}}_k}} $ and ${{\rm{vec}}({\bf{S}}^T)} = \sum\nolimits_{k = 1}^r {{{\left( {{\sigma _k} - \tau } \right)}_ + }{{\bf{u}}_k} \otimes {{\bf{v}}_k}} $,
      	which vary as the quantum states $\left| {{\psi _{{{\bf{A}}_0}}}} \right\rangle = 1/\sqrt{N_1} \sum\nolimits_{k = 1}^r {{\sigma _k}\left| {{{\bf{u}}_k}} \right\rangle \left| {{{\bf{v}}_k}} \right\rangle } $ and $\left| {{\psi _{\bf{S}}}} \right\rangle = 1/\sqrt{N_2}\sum\nolimits_{k = 1}^r {{{\left( {{\sigma _k} - \tau } \right)}_ + }\left| {{{\bf{u}}_k}} \right\rangle \left| {{{\bf{v}}_k}} \right\rangle }$ respectively, where $N_1 = \sum\nolimits_{k = 1}^r {\sigma _k^2}$ and $N_2 = \sum\nolimits_{k = 1}^r {{\left( {{\sigma _k} - \tau } \right)}_ +^2 }$.
     	Therefore, the QSVT algorithm solves the transformed problem ${\rm{vec}}({{\bf{S}}^T}) = {{\cal D}_\tau }\left( {{\rm{vec}}({\bf{A}}_0^T)} \right) $ \cite{Duan17}.

\subsection{QSVT algorithm}	
\label{sec2.2:level2}

     	Let ${\bf{A}} = {\bf{A}}_0{{\bf{A}}_0^\dag }$, therefore ${\bf{A}} = \sum\nolimits_{k = 1}^r {{\sigma _k^2}{{\bf{u}}_k}{\bf{u}}_k^T}$.
     	The QSVT algorithm is now shown as follows \cite{Duan17}:

     	\emph{Input.} A quantum state $\left| \psi_{{\bf{A}}_0} \right\rangle$, a unitary ${e^{ i{{{\bf{A} }}}{t_0}}}$, and a constant $\tau$.

     	\emph{Output.} A quantum state $\left| {{\psi _{\bf{S}}}} \right\rangle  $.

     	\emph{Algorithm.} $ {\bf{S}} = QSVT\left( { {\bf{A}}_0, \tau } \right)$.
     	The procedure of the algorithm can be illustrated as a sequence of the unitary operations:
                	\begin{equation}
                    \begin{aligned}
                        {{\bf{U}}_{QSVT}} = \left( {{{\bf{I}}^a} \otimes {\bf{U}}_{PE}^\dag } \right)\left( {{\bf{U}}_{cR} \otimes {{\bf{I}}^B}} \right)\left( {{{\bf{I}}^a} \otimes {{\bf{U}}_{PE}}} \right),
                    \end{aligned}
                    \label{eq:U_QSVT1}
                	\end{equation}
    	where ${\bf{U}}_{PE}$ and ${\bf{U}}_{cR}$ are the unitary operations of `phase estimation' and `controlled rotation', which are shown in Eqs. (\ref{eq:U_PE}) and (\ref{eq:Rf}), respectively, and ${\bf{U}}_{PE}^\dag$ represents the inverse of ${\bf{U}}_{PE}$.

    	Eq. (\ref{eq:U_QSVT1}) shows that the QSVT algorithm consists of two core subroutines, namely, ${\bf{U}}_{PE}$ and ${\bf{U}}_{cR}$.

    	(i) The first core subroutine ${{\bf{U}}_{PE}}$ is presented as follows:
                	\begin{equation}
                        \begin{aligned}
                            {{\bf{U}}_{PE}} = {{\bf{U}}_{PE}}\left( {\bf{A}} \right) 
                            = \left( {{\bf{F}}_{{\bf{T}}}^\dag \otimes {{\bf{I}}^B}} \right)\left( {\sum\nolimits_{\tau  = 0}^{T - 1} {\left| \tau  \right\rangle {{\left\langle \tau  \right|}^C} \otimes {e^{i{\bf{A}}\tau {{{t_0}} \mathord{\left/
                             {\vphantom {{{t_0}} T}} \right.
                             \kern-\nulldelimiterspace} T}}}} } \right)\left( {{{\bf{H}}^{ \otimes t}} \otimes {{\bf{I}}^B}} \right),
                        \end{aligned}
                        \label{eq:U_PE}
                	\end{equation}
    	where register $C$ stores the estimated eigenvalues of an Hermite matrix $\bf{A}$,
    	register $B$ stores the input state $\left| \psi_{{\bf{A}}_0} \right \rangle$,
    	${\bf{F}}_{{\bf{T}}}^\dag$ is the inverse quantum Fourier transform,
    	and ${\sum\nolimits_{\tau  = 0}^{T - 1} {\left| \tau  \right\rangle {{\left\langle \tau  \right|}^C} \otimes {e^{i{\bf{A}}\tau {{{t_0}} \mathord{\left/ {\vphantom {{{t_0}} T}} \right. \kern-\nulldelimiterspace} T}}}} }$ is the conditional Hamiltonian evolution \cite{HHL09}.
	The Hermite matrix ${\bf{A}}$ determines which eigenspace the quantum algorithm implements on.
    	Note that ${\bf{A}} = {\bf{A}}_0{{\bf{A}}_0^\dag }$, where ${{\bf{u}}_k}$ are the eigenvectors of ${\bf{A}}$ and the corresponding eigenvalues are ${\lambda _k} = \sigma _k^2$.
    	By taking a partial trace of $\left| {{\psi _{{\bf{A}}_0}}} \right\rangle \left\langle {{\psi _{{\bf{A}}_0}}} \right|$,
    	the density matrix that represents  ${\bf{A}}$  can be obtained \cite{SSP16b}:
	$ t{r_2}\left( \left| {{\psi _{{\bf{A}}_0}}} \right\rangle \left\langle {{\psi _{{\bf{A}}_0}}} \right| \right)
                = 1/{N_1}\sum\nolimits_{k = 1}^r {{\sigma _k^2}\left| {{{\bf{u}}_k}} \right\rangle \left\langle {{{\bf{u}}_k}} \right| = {\bf{A}} /tr{\bf{A}}} $.

    	(ii) The second core subroutine ${\bf{U}}_{cR}$ aims to `extract' and then  `reassign' the proportion of each eigenstate in the superposition $\left| \psi_{{\bf{A}}_0} \right\rangle = 1/\sqrt{N_1} \sum\nolimits_{k = 1}^r {{\sigma _k}\left| {{{\bf{u}}_k}} \right\rangle \left| {{{\bf{v}}_k}} \right\rangle }$.
	In particular, ${\bf{U}}_{cR}$ helps change the probability amplitude of each basic state $\left| {{{\bf{u}}_k}} \right\rangle \left| {{{\bf{v}}_k}} \right\rangle$ from ${\sigma _k}$ to $({\sigma _k}-\tau)_+$ via a transformation $({\sigma _k}-\tau)_+={\sigma _k} \times \frac{\left(\sqrt{\sigma _k^2}-\tau \right)_+}{\sqrt{\sigma _k^2}}$.
    Without loss of generality, ${\bf{U}}_{cR}$ is defined as follows: if $\sqrt{z} > \tau $:
        \begin{equation}
        \begin{aligned}
            \left| 0 \right\rangle \left| z \right\rangle 
            \to \left( {\frac{{\gamma \left( {\sqrt z  - \tau } \right)}}{{\sqrt z }}\left| 1 \right\rangle  + \sqrt {1 - \frac{{{\gamma ^2}{{\left( {\sqrt z  - \tau } \right)}^2}}}{z}} \left| 0 \right\rangle } \right) \left| z \right\rangle;
        \end{aligned}
        \label{eq:Rf}
        \end{equation}
     otherwise do nothing.

\section{QSVT circuit}
\label{sec3:level1}

	In this section, we further study the QSVT algorithm \cite{Duan17} based on the quantum circuit model. It provides the ability for the quantum computer to solve many NNM-based problems.
	Firstly, we describe the overview model of quantum circuit for QSVT. Secondly, we investigate in depth the realization of controlled rotation via quantum circuit
which involves the computation of ${\bf{U}}_{\sigma,\tau}$ and ${\bf{R}}_y$.
In the stage of ${\bf{U}}_{\sigma,\tau}$, we introduce Newton's method and simplify the Newton iteration function in terms of an intermediate variable ${z_k}$. We also reduce the number of Newton iterations with the help of a magic number $R$. In the stage of ${\bf{R}}_y$, we introduce an adjustable parameter $\alpha$, and demonstrate that the value of  $\alpha$ can be computed ahead of implementing the quantum circuit to ensure high probability and high fidelity readout.

\begin{figure}[ht]
\centering
\includegraphics[width=0.8\linewidth]{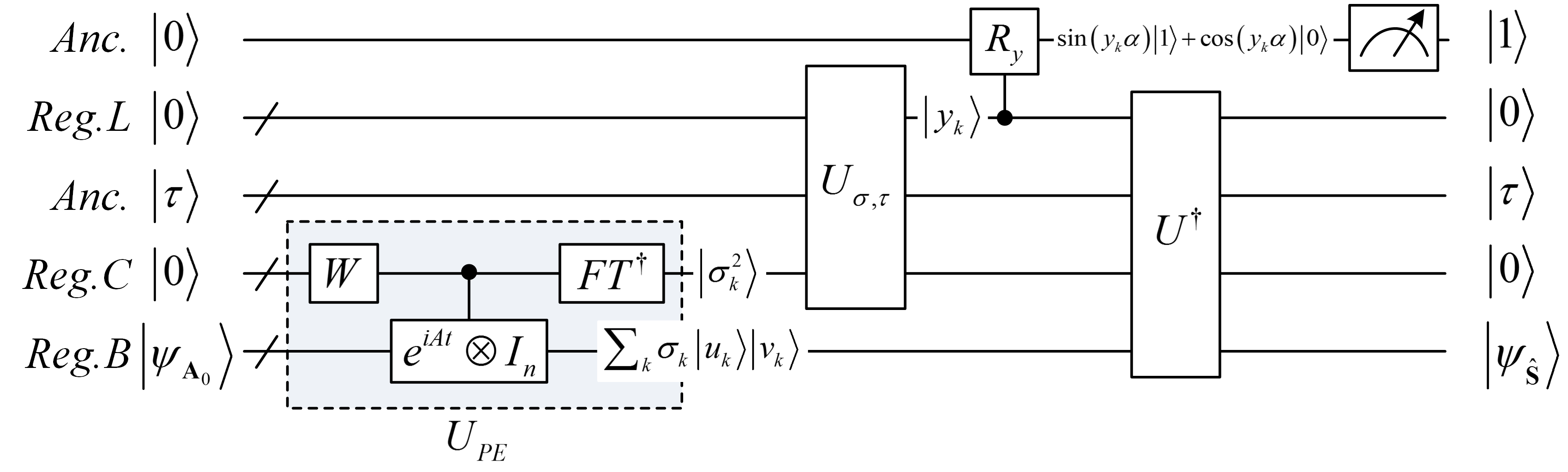}
\caption{Overview of the quantum circuit for solving the transformed SVT. Wires with `/' represent the groups of qubits. The norms of the quantum states are omitted for convenience.}
\label{fig1}
\end{figure}

    	The overview of the circuit for solving QSVT is shown in Fig. \ref{fig1}.
A solution of QSVT with error $O\left( \varepsilon  \right)$ can then be obtained.
    We omit the ancilla  $\left| \tau \right\rangle$ in the following register presentation because it remains the same during the procedure of the quantum circuit.
The detailed quantum circuit is presented as follows.

(1) Prepare the quantum registers in the state
        \begin{equation}
        \begin{aligned}
            \left| {{\psi _1 }} \right\rangle  = \left| 0 \right\rangle {\left| 0 \right\rangle ^L}{\left| 0 \right\rangle ^C}{\left| \psi_{{\bf{A}}_0}  \right\rangle ^B},
        \end{aligned}
        \label{eq_psi1}
        \end{equation}
        where the input matrix ${\bf{A}}_0$ has been prepared quantum mechanically as a quantum state $\left| {{\psi _{\bf{A_0}}}} \right\rangle$ stored in the quantum register $B$.
The number of qubits in the register $B$ is $b = O\left[ {{{\log }_2}\left( {pq} \right)} \right]$.

(2)  Perform the unitary operation ${{\bf{U}}_{PE}}( \bf{ A} )$ on the state $\left| 0 \right\rangle^C{\left| \psi_{{\bf{A}}_0}  \right\rangle ^B}$.
    	Recall $\left| {{\psi _{{{\bf{A}}_0}}}} \right\rangle  = 1/\sqrt{N_1} \sum\nolimits_{k = 1}^r {{\sigma _k}\left| {{{\bf{u}}_k}} \right\rangle \left| {{{\bf{v}}_k}} \right\rangle } $ where $\left| {{{\bf{u}}_k}} \right\rangle$ are the eigenstates of ${\bf{ A}}$ and $ \sigma_k^2$ are the corresponding eigenvalues.
	As shown in Ref. \cite{Duan17}, we have the state
        \begin{equation}
        \begin{aligned}
            \left| {{\psi _2}} \right\rangle  = \frac {1} {\sqrt{N_1}} {\left| 0 \right\rangle {\left| 0 \right\rangle ^L} }\sum\limits_{k = 1}^{r} {{\sigma _k}\left| { \sigma _k^2} \right\rangle ^C \left| {{{\bf{u}}_k}} \right\rangle \left| {{{\bf{v}}_k}} \right\rangle ^B}.
        \end{aligned}
        \label{eq_psi2}
        \end{equation}
Let the efficient condition number of ${\bf{ A}} $ be $\kappa$, such that ${\sigma _k}^2 \in \left[ {1/\kappa ,1} \right]$. The number of qubits in the register $C$ is $n = O\left( {{{\log }_2}\kappa } \right) $.

(3)   The unitary ${\bf{U}}_{\sigma,\tau}$ converts the eigenvalues $\left|  \sigma_k^2 \right\rangle$ stored in the register $C$ to the intermediate result $\left| { y}_k \right\rangle $ stored in the register $L$, where $ {{y}_k}   = \left(1- {\tau} / { \sigma_k}\right)_+ \in [0,1)$ . The state is

            \begin{equation}
            \begin{aligned}
                \left| {{\psi _3}} \right\rangle  = \frac {1} {\sqrt{N_1}} \left| 0 \right\rangle \sum\limits_{k = 1}^{r} {\sigma _k}  { {\left|  y_k \right\rangle ^L}}{\left| { \sigma _k^2} \right\rangle ^C \left| {{{\bf{u}}_k}} \right\rangle \left| {{{\bf{v}}_k}} \right\rangle ^B}.            			\end{aligned}
                \label{eq_psi3}
            \end{equation}
	According to the ideas of Ref. \cite{CPP12},  Newton iteration can be used to realize ${\bf{U}}_{\sigma,\tau}$ (see Sec. \ref{sec:U}).
The number of qubits in the register $L$ and the ancilla  $\left| \tau \right\rangle$ are both $d = O\left( {{{\log }_2}\kappa } \right)$.

(4)   To realize   ${\bf{R}}$:        	
	$ \left| 0 \right\rangle \left| y_k \right\rangle   \mapsto  \left( {y_k\left| 1 \right\rangle  + \sqrt {1 - {y_k^2}} \left| 0 \right\rangle } \right)\left| y_k \right\rangle$,
	 i.e., to  `extract' the value of $y_k$ in register $L$ to the amplitude of the ancilla qubit,  we introduce a unitary ${\bf{R}}_y$ with parameter $\alpha$ (see Sec. \ref{sec:R}) to approximate ${\bf{R}}$:

         \begin{equation}
            	{\bf{R}_y}\left| 0 \right\rangle \left| y_k \right\rangle  = \left[ {  \sin(y_k\alpha)  \left| 1 \right\rangle  + \cos(y_k\alpha) \left| 0 \right\rangle } \right]\left| y_k \right\rangle ,0 < y_k < 1.
            	 \label{R_y}
         \end{equation}

        Subsequently, ${\bf{R}}_y$  is applied to the ancilla qubit on the top of the circuit and controlled by the register $L$. We obtain the state

            \begin{equation}
            \begin{aligned}
                \left| {{\psi _4}} \right\rangle  = \frac {1} {\sqrt{N_1}} \sum\limits_{k = 1}^{r} {\sigma _k}       \left[     {{\sin \left(  y_k \alpha \right)} {\left| 1 \right\rangle }  + {\cos \left(  y_k \alpha \right)} {\left| 0 \right\rangle }  }   \right]     { {\left|  y_k \right\rangle ^L}}{\left| { \sigma _k^2} \right\rangle ^C \left| {{{\bf{u}}_k}} \right\rangle \left| {{{\bf{v}}_k}} \right\rangle ^B}.            \end{aligned}
                \label{eq_psi4}
            \end{equation}

(5)  Uncompute the registers $L$, $C$ and $B$, and remove the register $L$ and $C$, we have
            \begin{equation}
            \begin{aligned}
                \left| {{\psi _5}} \right\rangle  = \frac {1} {\sqrt{N_1}} \sum\limits_{k = 1}^{r} {\sigma _k}       \left[     {{\sin \left(  y_k \alpha \right)} {\left| 1 \right\rangle }  + {\cos \left(  y_k \alpha \right)} {\left| 0 \right\rangle }  }   \right]     { \left| {{{\bf{u}}_k}} \right\rangle \left| {{{\bf{v}}_k}} \right\rangle ^B}.
            \end{aligned}
            \label{eq_psi5}
            \end{equation}

(6)  Measure the top ancilla bit. If the result returns 1, then the register $B$ of the system collapses to the state final state
            \begin{equation}
            \begin{aligned}
                \left| \psi _{\bf{\hat  S}} \right\rangle  = \frac {1} {\sqrt{N_\alpha}} \sum\limits_{k = 1}^{r} {\sigma _k}     {\sin \left(  y_k \alpha \right)}  { \left| {{{\bf{u}}_k}} \right\rangle \left| {{{\bf{v}}_k}} \right\rangle ^B},
            \end{aligned}
            \label{eq_psi6}
            \end{equation}
	where $N_\alpha =  \sum\nolimits_{k = 1}^{r} {\sigma _k}^2     {\sin ^2 \left(  y_k \alpha \right)}  $.

    In summary, the procedure of the quantum circuit can be illustrated as follows:
                	\begin{equation}
                    \begin{aligned}
                        {{ \bf{\tilde U}}_{QSVT}} =
                        \left( {{\bf{I}}^{aLa} \otimes {{\bf{U}}}_{PE}^\dag } \right)
                        \left( {{\bf{I}}^{a} \otimes {\bf{U}_{\sigma,\tau}^\dag} \otimes {\bf{I}}^{B}} \right)
                        \left( {{\bf{R}}_y \otimes {\bf{I}}^{aCB}} \right)
                        \left( {{\bf{I}}^{a} \otimes {\bf{U}_{\sigma,\tau}} \otimes {\bf{I}}^{B}} \right)
                        \left( {{\bf{I}}^{aLa} \otimes {\bf{U}}_{PE}} \right).
                    \end{aligned}
                    \label{eq:U_QSVT2}
                	\end{equation}

\subsection { Computation of ${\bf{U}}_{\sigma,\tau}$}
\label{sec:U}

\begin{figure}[ht]
        \centering
        \includegraphics[width=0.9\linewidth]{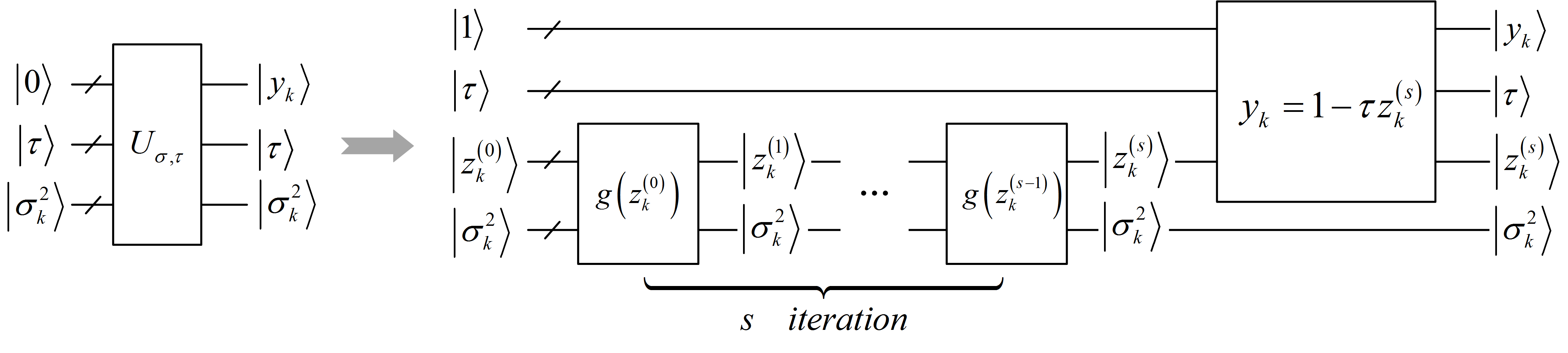}
        \caption{A quantum circuit of the unitary ${\bf{U}}_{\sigma,\tau}$.}
        \label{figUsigmatau}
        \end{figure}

        We now deal with the detailed quantum circuit of the unitary ${\bf{U}}_{\sigma,\tau}$, that is, we deal with the function of the eigenvalues ${\sigma _k^2}$ of  $\bf{A}$ : $ {{y}_k}   = \left(1- {\tau} / {{\sqrt {\sigma _k^2} }}\right)_+ \in [0,1)$. Here Newton iteration is introduced to compute $y_k = 1 - {\tau }/{{\sqrt {\sigma _k^2} }}$ for ${\sigma _k} > \tau $. For simplifying the quantum circuit design of ${\bf{U}}_{\sigma,\tau}$,  an intermediate variable ${z_k}$ is introduced such that ${z_k} = z\left( {\sigma _k^2} \right) = {1/{{\sqrt {\sigma _k^2} }}}$ and ${y_k} = y\left( {{z_k}} \right) = 1 - \tau {z_k}$. Therefore, Newton iteration can be just used to compute the simplified function ${z_k} = z\left( {\sigma _k^2} \right) = {1/{{\sqrt {\sigma _k^2} }}}$. From  above, a quantum circuit of the unitary ${\bf{U}}_{\sigma,\tau}$ can be designed as shown in Fig. \ref{figUsigmatau}.

        Specifically, the detailed quantum circuit design of ${\bf{U}}_{\sigma,\tau}$ can be divided into two parts. The first part is the rightmost unitary  ${y_k} = 1 - \tau {z_k}$ as shown in Fig. \ref{figUsigmatau}, and the second part is the leftmost Newton iterations.

\begin{figure}[ht]
        \centering
        \includegraphics[width=0.8\linewidth]{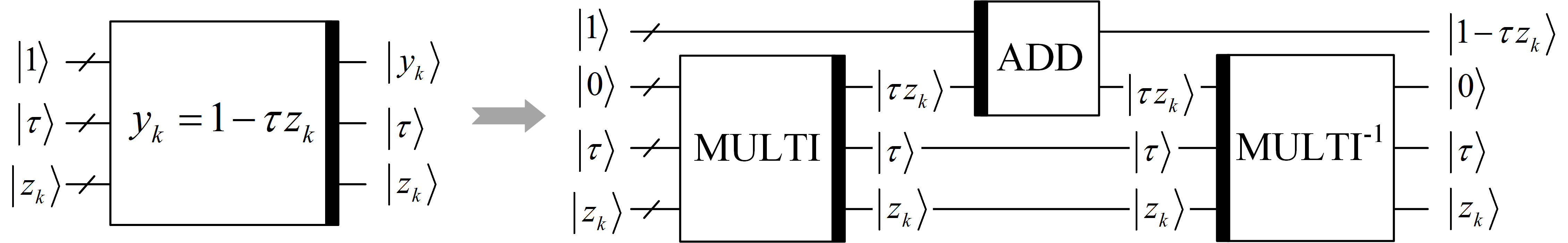}
        \caption{Quantum circuit for computing ${y_k} = 1 - \tau {z_k}$. The black strip in the right side of the rectangle represents a unitary operation, and the black strip in the left one represents the inverse of the corresponding unitary operation.}
        \label{figyk}
        \end{figure}

        Obviously, the first part can be simply realized via the quantum circuits for addition and multiplication which have been studied in Refs. \cite{VBE96,PRM17}.
        The corresponding quantum circuit is shown in Fig. \ref{figyk}.
        The number of qubits is $O\left(d\right)$ and the number of quantum operations for implementing addition and multiplication is $O\left(d^2\right)$.

        Let us turn to the second part. A method for solving $z\left( {\sigma _k^2} \right) = {1/{{\sqrt {\sigma _k^2} }}}$ has been presented in Ref. \cite{BHP15}. It provides the quantum circuit for the initial state $\left| {z_k^{\left( 0 \right)}} \right\rangle  = \left| {{2^{\left\lfloor {\frac{{w - 1}}{2}} \right\rfloor }}} \right\rangle $, where $w \in N$ and ${2^{1 - w}} > x > {2^{ - w}}$, and the idea for solving iteration function in terms of an abstraction of an elementary module. Here we make further research of the Newton method. Concretely, we give the detailed quantum circuit of the Newton iteration in terms of basic elementary gates, and introduce a different initial state which helps to reduce the iteration steps.

         Applying the Newton method to the function $f\left( {{z_k}} \right) = {1 /{z_k^2}} - \sigma _k^2$, we can obtain the iteration function

       \begin{equation}
       \begin{aligned}
            z_k^{\left( {i + 1} \right)} &= g\left( {z_k^{\left( i \right)}} \right) = z_k^{\left( i \right)} - \frac{{f\left( {z_k^{\left( i \right)}} \right)}}{{f'\left( {z_k^{\left( i \right)}} \right)}} = z_k^{\left( i \right)} - \frac{{{{{\left( {z_k^{\left( i \right)}} \right)}^{-2}}} - \sigma _k^2}}{{ - 2{{\left( {z_k^{\left( i \right)}} \right)}^{ - 3}}}}\\
             &= \frac{1}{2}\left( {3z_k^{\left( i \right)} - \sigma _k^2{{\left( {z_k^{\left( i \right)}} \right)}^3}} \right)
       \label{eq:gz}
       \end{aligned}
       \end{equation}
       where $i = 1,2,...,s$.

\begin{figure}[ht]
        \centering
        \includegraphics[width=1\linewidth]{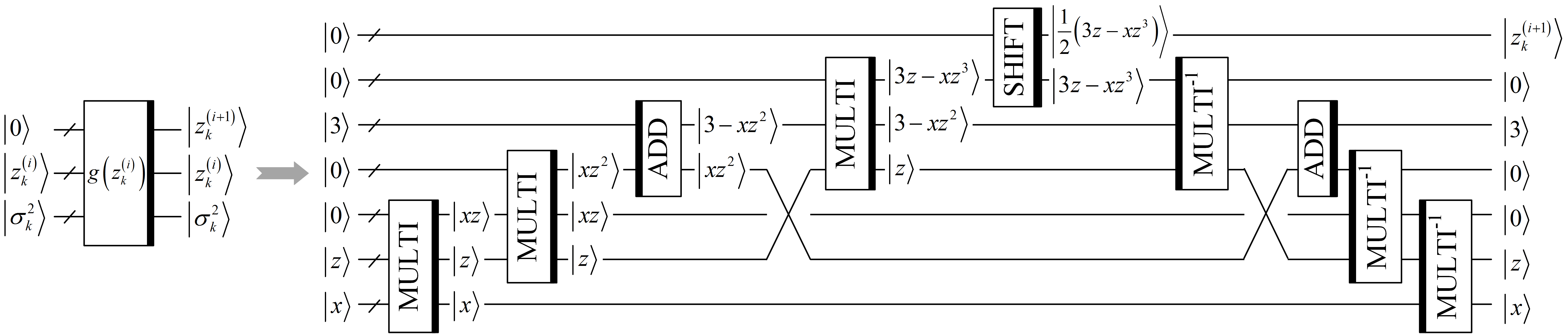}
        \caption{A quantum circuit of one Newton iteration for computing $g\left( {z_k^{\left( i \right)}} \right)$.}
        \label{figUgi}
        \end{figure}

        The detailed quantum circuit of $g\left( {z_k^{\left( i \right)}} \right)$ is presented in Fig. \ref{figUgi}.
        Inevitably, four extra quantum registers are needed for the inverse of the unitary operators, and one more ancilla register is needed for storing the output $z_k^{\left( {i + 1} \right)}$ in each iteration $g\left( {z_k^{\left( i \right)}} \right)$.
        As the circuit of the iteration is composed of basic operations ( i.e. addition, multiplication and shift operations), each iterative step requires $O\left( {n + d} \right)$ qubits and $O\left[ {ploy\left( {n + d} \right)} \right]$ quantum operations, where the degree of the polynomial is no more than $3$ according to Refs. \cite{VBE96,PRM17}.

\begin{figure}[ht]
        \centering
        \includegraphics[width=0.9\linewidth]{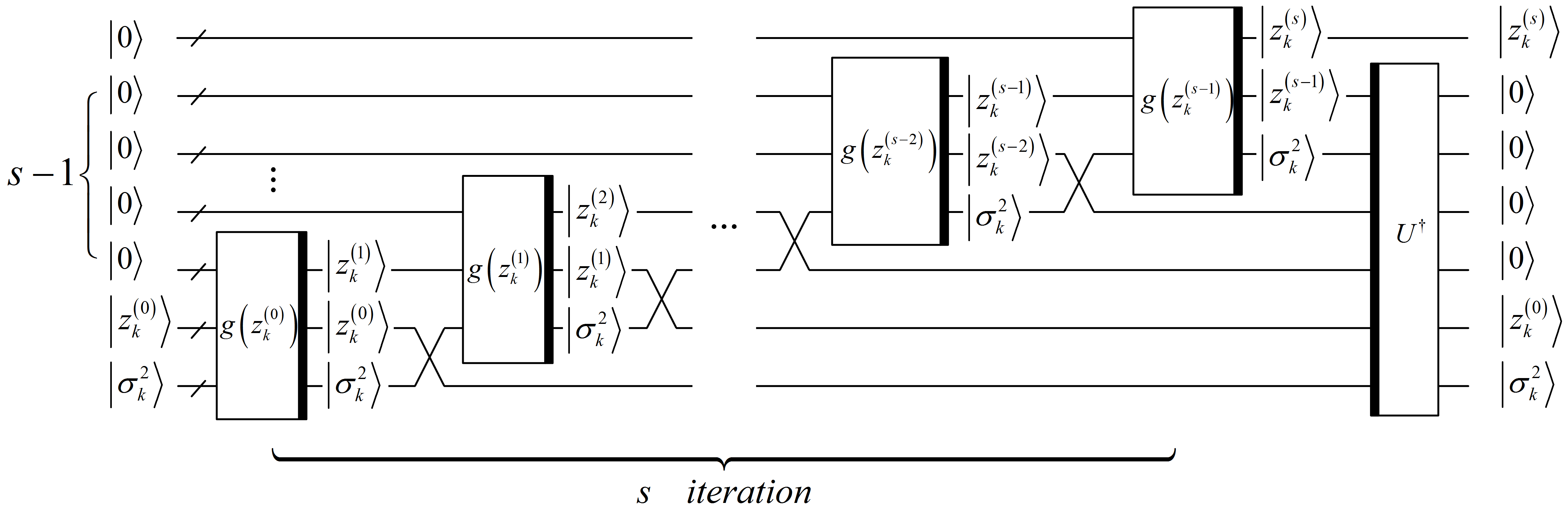}
        \caption{A quantum circuit for computing $ {z_k^{\left(s\right)}} $ with $s$ iterations.}
        \label{figUg}
        \end{figure}

        With indispensable ancilla registers, a quantum circuit for computing $ {z_k^{\left(s\right)}} $ after $s$ Newton iterations is designed as in Fig. \ref{figUg}.
        Therefore, the number of qubits for implementing the Newton iteration is $O\left( {n + sd} \right)$, and the number of quantum operations is $O\left[ {s \cdot ploy\left( {n + d} \right)} \right]$.

        As in Fig. \ref{figUg}, after each iteration, at least one extra register is needed for storing the intermediate output $z_k^{\left(i\right)}$. Therefore, after $s$ iterations, $s-1$ garbage registers are produced.

\begin{figure}[ht]
        \centering
        \includegraphics[width=0.4\linewidth]{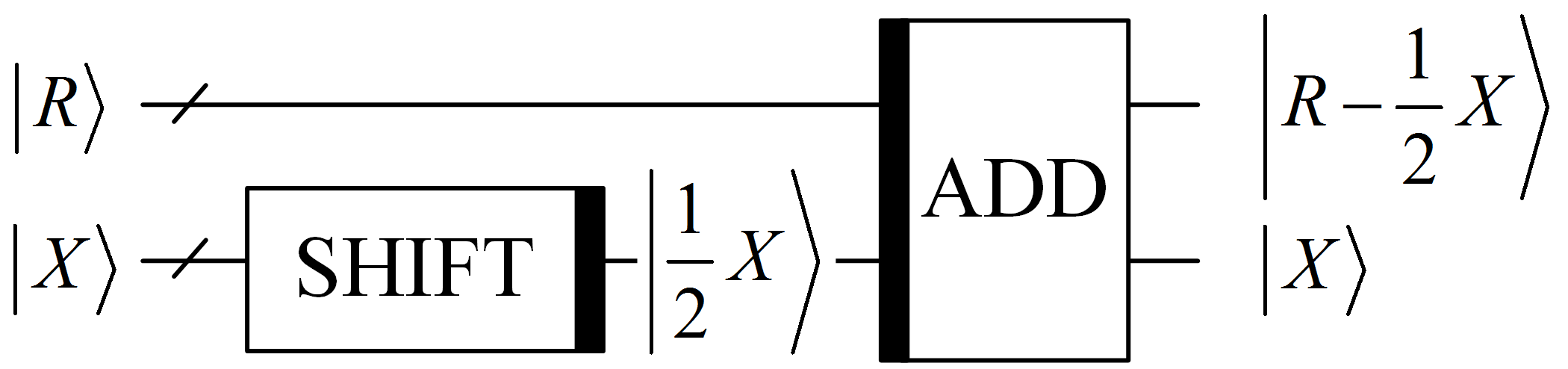}
        \caption{A quantum circuit for computing the initial state $z_k^{\left(0\right)}$.} $X$ is presented as initial $z_k^{\left(0\right)}$ which follows the IEEE 754 floating-point format.
        \label{figz0}
        \end{figure}

        To reduce the number of iterations, we choose a magic number $R$ in Ref. \cite{INVSQRT} rather than the initial state in Ref. \cite{BHP15}. The magic number $R$ produces the first approximation of the initial state by $X = R - \left( {X >  > 1} \right)$ which follows the IEEE 754 floating-point format \cite{IEEE754}. It helps Newton's method run only one or two iterations and output a more precise approximation (see Appendix \ref{A0}). Then the number of qubits for implementing the Newton iteration can be reduced to $O\left( {n + d} \right)$, and the number of quantum operations can be reduced to $O\left[ { ploy\left( {n + d} \right)} \right]$. The corresponding quantum circuit is shown in Fig. \ref{figz0}.

         Now turning to the error analysis of the quantum cicuit of the unitary ${\bf{U}}_{\sigma,\tau}$. It consists of two parts. The first is the error $e_s$ caused by the Newton's iteration, the second is the roundoff error ${{\hat e}_s}$ caused by truncating the result of each iteration to $d$ qubits of accuracy before passing it to the next iteration.

         According to the Eq. (\ref{eq:gz}), we have
          \begin{equation}
           \begin{aligned}
                g\left( {z_k^{\left( i \right)}} \right) - \frac{1}{{\sqrt {\sigma _k^2} }} &= \frac{1}{2}\left( {3z_k^{\left( i \right)} - \sigma _k^2{{\left( {z_k^{\left( i \right)}} \right)}^3}} \right) - \frac{1}{{\sqrt {\sigma _k^2} }} \\
                &=  - \frac{1}{2}{\left( {z_k^{\left( i \right)} - \frac{1}{{\sqrt {\sigma _k^2} }}} \right)^2}\sqrt {\sigma _k^2} \left( {z_k^{\left( i \right)}\sqrt {\sigma _k^2}  + 2} \right)
           \end{aligned}
           \end{equation}
       The last quantity is non-positive for ${z_k^{\left( i \right)}} \ge 0$. Similar to the equation in Ref. \cite{BHP15}, the error $e_s$ satisfies
           \begin{equation}
           \begin{aligned}
            {e_s}&:= \left| {z_k^{\left( s \right)} - \frac{1}{{\sqrt {\sigma _k^2} }}} \right| = \frac{1}{2}e_{s - 1}^2\sqrt {\sigma _k^2} \left( {z_k^{\left( {s - 1} \right)}\sqrt {\sigma _k^2}  + 2} \right)\\
             & \le \frac{3}{2}\sqrt {\sigma _k^2} e_{s - 1}^2 \le \frac{2}{{3\sqrt {\sigma _k^2} }}{\left( {\frac{3}{2}\sqrt {\sigma _k^2} {e_0}} \right)^{{2^s}}}
           \end{aligned}
           \end{equation}
       The initial error ${e_0}$  satisfies  $\sqrt {\sigma _k^2} {e_0} \le 1/8$ for the initial state derived from the magic number $R$ \cite{INVSQRT}, which is better than $\sqrt {\sigma _k^2} {e_0} \le 1/2$ for the traditional initial state $\left| {z_k^{\left( 0 \right)}} \right\rangle  = \left| {{2^{\left\lfloor {\frac{{w - 1}}{2}} \right\rfloor }}} \right\rangle $ \cite{BHP15}.

       As shown in Ref. \cite{BHP15}, the truncation error ${{\hat e}_s}$ satisfies
           \begin{equation}
           \begin{aligned}
                {{\hat e}_s}: = \left| {{{\hat z}^{\left( s \right)}} - {z^{\left( s \right)}}} \right| \le {2^{1 - d}} \cdot {\left( {\frac{3}{2}} \right)^s}
           \end{aligned}
           \end{equation}

       To sum up, with the initial state in \cite{BHP15}, the number of the iteration steps is $s = O\left( {{{\log }_2}d} \right)$. Then the error caused by the unitary ${\bf{U}}_{\sigma,\tau}$ is
           \begin{equation}
           \begin{aligned}
                \left| {z_k^{\left( s \right)} - \frac{1}{{\sqrt {\sigma _k^2} }}} \right| &\le {e_s} + {{\hat e}_s} \le \frac{2}{{3\sqrt {\sigma _k^2} }}{\left( {\frac{3}{4}} \right)^{{2^s}}} + {2^{1 - d}} \cdot {\left( {\frac{3}{2}} \right)^s}\\
                 &\le \frac{2}{{3\sqrt {\sigma _k^2} }}{\left( {\frac{3}{4}} \right)^d} + {2^{1 - d}} \cdot {2^{{{\log }_2}d + 1}}\\
                 &\le \frac{{2\sqrt \kappa  }}{3}{\left( {\frac{3}{4}} \right)^d} + {2^{2 - d}} \cdot d
           \end{aligned}
           \end{equation}

       And with the initial state in \cite{INVSQRT}, the number of the iteration steps is $s=1$ or $s=2$. Then the error caused by the unitary ${\bf{U}}_{\sigma,\tau}$ are
           \begin{equation}
           \begin{aligned}
                \left| {z_k^{\left( 1 \right)} - \frac{1}{{\sqrt {\sigma _k^2} }}} \right| \le {e_1} + {{\hat e}_1} \le \frac{2}{{3\sqrt {\sigma _k^2} }}{\left( {\frac{3}{{16}}} \right)^2} + {2^{1 - d}} \cdot \left( {\frac{3}{2}} \right)
                 \le \frac{{3\sqrt \kappa  }}{{{2^7}}} + \frac{3}{{{2^d}}}
           \end{aligned}
           \end{equation}
       and
            \begin{equation}
           \begin{aligned}
                \left| {z_k^{\left( 2 \right)} - \frac{1}{{\sqrt {\sigma _k^2} }}} \right| \le {e_2} + {{\hat e}_2} \le \frac{2}{{3\sqrt {\sigma _k^2} }}{\left( {\frac{3}{{16}}} \right)^4} + {2^{1 - d}} \cdot {\left( {\frac{3}{2}} \right)^2} \le \frac{{27\sqrt \kappa  }}{{{2^{15}}}} + \frac{9}{{{2^{d + 1}}}}
           \end{aligned}
           \end{equation}
       respectively.

\subsection { Computation of ${\bf{R}}_y$}
\label{sec:R}

\begin{figure}[ht]
\centering
\includegraphics[width=0.6\linewidth]{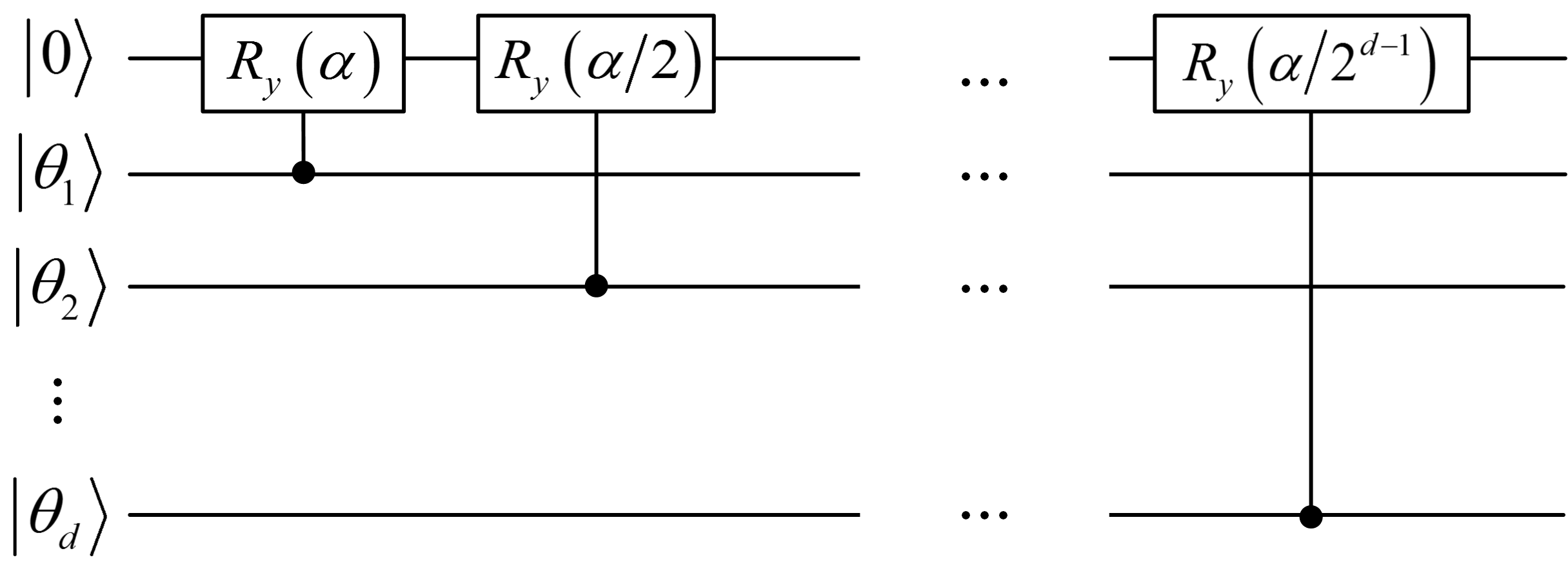}
\caption{Quantum circuit of the unitary ${\bf{R}}_y$.}
\label{fig4}
\end{figure}

    	A sequence of rotations about the $y$ axis can be used to implement the unitary ${\bf{R}}_y$ \cite{CPP12},
	where ${\bf{R}}_y(2\theta)\left| 0 \right\rangle = \sin \theta \left| 1 \right\rangle + \cos \theta \left| 0 \right\rangle $.	
	Here we introduce a parameter $\alpha$, which can be used to improve the probability and accuracy of obtaining the final state (see Sec. \ref{sec:PF}).
	Assume that we have obtained a $d$-qubit state $\left| y_k  \right\rangle   \buildrel \Delta \over  =  \left| \theta  \right\rangle $.
    	Consider the binary representation of $\theta$: $\theta  = 0.{\theta _1} \cdots {\theta _d} = \sum\nolimits_{j = 1}^d {{\theta _j} \cdot {2^{ - j}}} ,\left( {{\theta _j} \in \left\{ {0,1} \right\}} \right)$, there is
        \begin{equation}
            \begin{aligned}
                {{\bf{R}}_y}\left( {2\alpha\theta } \right)
                 = {e^{ - i\alpha\theta Y}} = \prod\limits_{j = 1}^d {{e^{ - iY\alpha{{{\theta _j}} \mathord{\left/
                 {\vphantom {{{\theta _j}} {{2^j}}}} \right.
                 \kern-\nulldelimiterspace} {{2^j}}}}}}
      		= \prod\limits_{j = 1}^d {{\bf{R}}_y^{{\theta _j}}\left( {{2^{1 - j}}}\alpha \right)}
            		= R_y^{{\theta _1}}\left( \alpha  \right)R_y^{{\theta _2}}\left( {{\alpha  \mathord{\left/
             {\vphantom {\alpha  2}} \right.
             \kern-\nulldelimiterspace} 2}} \right)
              \cdots R_y^{{\theta _d}}\left( {{\alpha  \mathord{\left/
             {\vphantom {\alpha  {{2^{d - 1}}}}} \right.
             \kern-\nulldelimiterspace} {{2^{d - 1}}}}} \right).
            \end{aligned}
        \label{R_y}
        \end{equation}
    where $Y$ is the Pauli $Y$ operator. The quantum circuit of  ${\bf{R}}_y$ is shown in Fig. \ref{fig4}.

{\subsection {Complexity analysis}}

    We now analyze the space and time resources required for the whole quantum circuit. The numbers of qubits in register $B$, $C$ and $L$ are $b = O\left[ {{{\log }_2}\left( {pq} \right)} \right]$, $n = O\left( {{{\log }_2}\kappa } \right) $ and $d = O\left( {{{\log }_2}\kappa } \right)$ respectively. Ancilla qubits in the computation of ${\bf{U}}_{\sigma,\tau}$ take up most space in the quantum circuit and the number of the ancilla qubits is $O {\left( {n + d} \right)}$. Typically, the condition number $\kappa$ is taken as $\kappa  = O\left( {1/\varepsilon } \right)$. Therefore, the total number of qubits required in the quantum circuit is $O\left[ {{{\log }_2}\left( {pq} /{\varepsilon} \right)} \right]$.

    Turing to the time cost. Phase estimation requires $O\left( {{n^2}} \right)$ operations and one call to the controlled-$\bf{A}$ black box \cite{NC10}. The number of quantum gates in the computation of ${\bf{U}}_{\sigma,\tau}$ is $O\left[ {ploy\left( {n + d} \right)} \right]$, and the number of gates in the computation of ${\bf{R}}_y$ is $O\left(d\right)$. Therefore, the total number of gates required in the quantum circuit is $O\left[ {ploy\log_2 \left( {1/\varepsilon } \right)} \right]$.

    In summary, the number of qubits required by the circuit is $O\left[ {{{\log }_2}\left( {pq} /{\varepsilon} \right)} \right]$ and the number of quantum operations used by the circuit is a low degree polynomial in $\log_2\left( {1/\varepsilon }\right)$.

\subsection {Probability and fidelity analysis}
\label{sec:PF}
	We now analyze the probability of obtaining the final result and the fidelity of the ideal and actual outputs.
	Eqs. (\ref{eq:U_QSVT1}) and (\ref{eq:U_QSVT2}) show that the quantum circuit ${{ \bf{\tilde U}}_{QSVT}}$ is a unitary approximation of the quantum algorithm ${{\bf{ U}}_{QSVT}}$.
    Without loss of generality, we assume that there is no error in any step of ${{\bf{ U}}_{QSVT}}$ and ${{ \bf{\tilde U}}_{QSVT}}$.
    Therefore, the unitary ${\bf{R}}_y$ in ${{ \bf{\tilde U}}_{QSVT}}$ dominates the probability and the fidelity in the quantum circuit.

	The probability of obtaining the final result  $\left| {{\psi _{\bf{\hat  S}}}} \right\rangle$ in Eq. (\ref{eq_psi6}) can be calculated via Eq. (\ref{eq_psi5}):
	
	    \begin{equation}
                P(\alpha)  = \frac {N_\alpha} {N_1} = \frac {\sum\nolimits_{k = 1}^{r} {\sigma _k}^2     {\sin ^2 \left(  y_k \alpha \right)}}   {\sum\nolimits_{k = 1}^{r} {\sigma _k}^2}.
            \label{eq_P}
            \end{equation}

	The fidelity of the actual and ideal final results can be calculated via the inner product of the output $\left| {{\psi _{\bf{\hat  S}}}} \right\rangle$ and the theoretical output $\left| {{\psi _{\bf{S}}}} \right\rangle = 1/\sqrt{N_2}\sum\nolimits_{k = 1}^r {{{\left( {{\sigma _k} - \tau } \right)}_ + }\left| {{{\bf{u}}_k}} \right\rangle \left| {{{\bf{v}}_k}} \right\rangle } = 1/\sqrt{N_2}\sum\nolimits_{k = 1}^r {{ {\sigma _k} y_k }\left| {{{\bf{u}}_k}} \right\rangle \left| {{{\bf{v}}_k}} \right\rangle }$:

	    \begin{equation}
	    \begin{aligned}
                    F(\alpha) &= \left\langle {{{\psi _{\hat S}}}}
                    {\left | {\vphantom {{{\psi _{\hat S}}} {{\psi _S}}}}
                    \right. }
                    {\psi _S} \right\rangle
                    = \frac{1}{\sqrt {{N_2}{N_\alpha}} } \sum\nolimits_{k = 1}^{r} {\sigma _k^2  y_k \sin \left( {y_k \alpha } \right)} \\
        		   &= \frac { \sum\nolimits_{k = 1}^r {\sigma _k^2  y_k \sin \left( {y_k \alpha } \right)} }   {\sqrt { {\sum\nolimits_{k = 1}^r {\sigma _k^2  y_k^2}} \times {\sum\nolimits_{k = 1}^{r} {\sigma _k}^2     {\sin ^2 \left(  y_k \alpha \right)} }   }}.
                    \label{eq_F}
            \end{aligned}
            \end{equation}

    Typically, $\alpha$ can be obtained by the method in Ref. \cite{CD16} such that ${y_k}\alpha   \approx  {\sin ^{ - 1}}\left( {{y_k}} \right)$. Therefore, the probability of obtaining the final result is

    	    \begin{equation}
                {P_{1}} \approx \frac{{\sum\nolimits_{k = 1}^r {{\sigma _k}^2} y_k^2}}{{\sum\nolimits_{k = 1}^r {{\sigma _k}^2} }}.
            \label{eq_P1}
            \end{equation}
    And the fidelity of the output approximates to 1. But the method in Ref. \cite{CD16} involves quantum circuits for exponentiation computations and the results in excessive time and space consumptions.

    Here, we introduce a method to compute $\alpha$ ahead of implementing the QSVT quantum circuit to avoid extra circuits involved in Ref. \cite{CD16}.
	To ensure high probability and high fidelity, we introduce function $G$ of $\alpha$:
	\begin{equation}
	\begin{aligned}
                   G(\alpha) &= \sqrt{P(\alpha)} \times {F(\alpha)} = \sqrt{\frac {N_\alpha} {N_1}} \times {\frac{1}{\sqrt {{N_2}{N_\alpha}} } \sum\nolimits_{k = 1}^{r} {\sigma _k^2  y_k \sin \left( {y_k \alpha } \right)} } \\
                   &= {\frac{\sum\nolimits_{k = 1}^{r} {\sigma _k^2  y_k \sin \left( {y_k \alpha } \right)}}{\sqrt {{\sum\nolimits_{k = 1}^{r} {\sigma _k}^2} \times {\sum\nolimits_{k = 1}^r {\sigma _k^2  y_k^2}} } }  },
         \label{eq_G}
       \end{aligned}
       \end{equation}
	and  to find $\alpha$ that maximizes $G(\alpha)$, i.e. to solve the problem $\mathop {\arg \max }\limits_\alpha  G\left( \alpha  \right)$. This problem is transformed to solve the following equation:
	
	\begin{equation}
		G'(\alpha) ={\frac{\sum\nolimits_{k = 1}^{r} {\sigma _k^2  y_k^2 \cos \left( {y_k \alpha } \right)}}{\sqrt {{\sum\nolimits_{k = 1}^{r} {\sigma _k}^2} \times {\sum\nolimits_{k = 1}^r {\sigma _k^2  y_k^2}} } }  }=0.
	\label{eq_G'}
    	\end{equation}
	
	A series of methods, such as gradient descent, Newton's method, evolutionary algorithms,  can be used to solve Eq. (\ref{eq_G'}). Given that the problem to be solved is not convex, these iterative algorithms can only output locally optimal solution.
	Taylor's series can also be used to solve this problem and obtain an approximate solution. 	
	
	Instead of using the aforementioned methods, we select an `intuitive' method to compute an approximate solution $\tilde \alpha$ for this problem.
   Recall that $y_k =  \left( 1-\tau/\sigma_k \right)_+$, we have $y_1 > y_2 \ge  \cdots \ge y_r$ as $\sigma_1 > \sigma_2 >  \cdots > \sigma_r > 0$.
	The  period of the $\cos\left(y_k\alpha \right)$ are $2\pi/y_k$, thereby satisfying $2\pi/y_1 < 2\pi/y_2 \leq  \cdots \leq 2\pi/y_r$. We now consider the case that the value $\alpha$ satisfies $y_1\alpha \in [0,\pi/2]$, therefore,
    $0 \leq \cos\left(y_1\alpha \right) < \cos\left(y_2\alpha \right) \leq  \cdots \leq \cos\left(y_r\alpha \right)$.
	
	Eq. (\ref{eq_G'}) shows that ${\sum\nolimits_{k = 1}^{r} {\sigma _k^2  y_k^2 \cos \left( {y_k \alpha } \right)}}=0$. Therefore,
	\begin{equation}	
	\begin{aligned}
                &\sigma _1^2y_1^2\cos \left( {{y_1}\alpha } \right) =  - \sigma _2^2y_2^2\cos \left( {{y_2}\alpha } \right) -  \cdots  - \sigma _r^2y_r^2\cos \left( {{y_r}\alpha } \right) \le 0 ,
	\end{aligned}
    	\end{equation}
	and considering $\cos\left(y_1\alpha \right) \in [0,1]$, we obtain the approximate solution $\tilde \alpha = \frac{\pi}{2y_1} = \frac{\pi}{2\left(1-\tau/\sigma_1 \right)}$.
	Note that $\tilde \alpha$ is close to be optimal when
${\sum\nolimits_{k = 1}^{r} {\sigma _k^2  y_k^2 \cos \left( {y_k \alpha } \right)}}$ is dominated by
$\sigma _1^2y_1^2\cos \left( {{y_1}\alpha } \right) $, because in this case the derivative approximates zero. That is, $\tilde \alpha$ is most effective when $\sigma_1$ dominates the most part of all singular values. 
	
	Although the approximate solution is not the optimal solution, it has two advantages. Firstly, it has simpler expression and can be computed more efficiently  compared with  iterative algorithms. Secondly, this `intuitive' solution is only based on the maximized singular value of ${\bf{A}}_0$ , i.e. priori knowledge of ${\bf{A}}_0$ is only the maximum singular value  $\sigma_1$ instead of all singular values of ${\bf{A}}_0$ compared with Taylor's series method (see Appendix \ref{A1}).
    In Appendix \ref{A2}, we make two experiments to analyze the probability and fidelity. The first experiment shows that $\tilde \alpha $ is likely to be a good solution to ensure high probability and fidelity. And the second experiment shows how the input matrix and the hyperparameter $\tau$ impact the probability and fidelity readout. 

\section {Example}
\label{sec4:level1}
    In this section, we design and implement a numerical simulation experiment of a small-scale QSVT circuit and analyze the results. The purpose for this example is to illustrate the algorithm and for potential experimental implementation using currently available resource.

\begin{figure}[ht]
\centering
\includegraphics[width=0.9\linewidth]{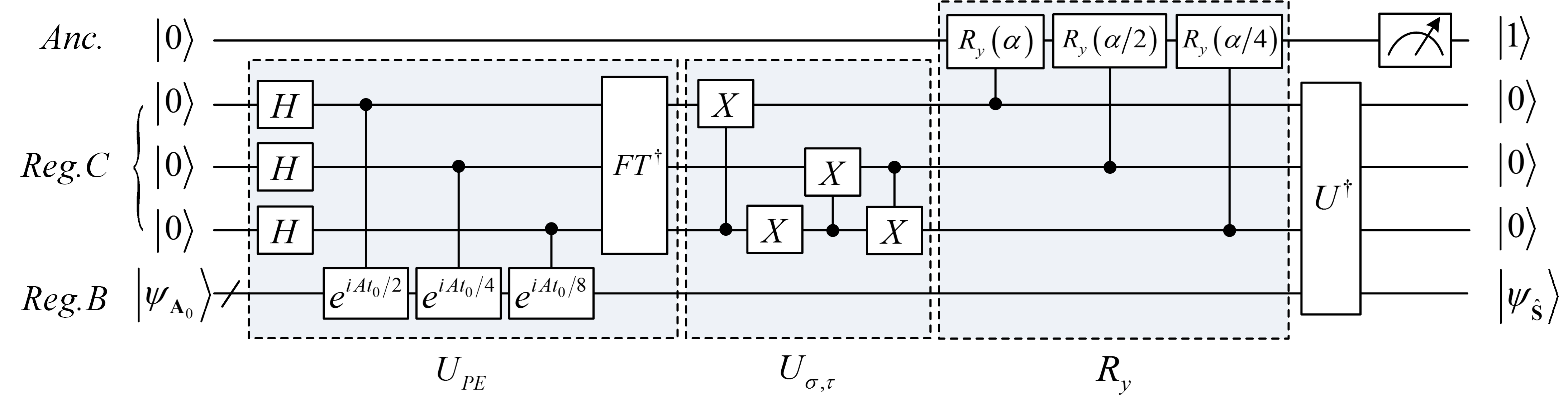}
\caption{Example quantum circuit for solving the SVT.
            ${{\bf{U}}^\dag }$ represents the inverse of all the operations before of ${\bf{R}}_y$. Each simplified unitary  $e^{i{\bf{A}}t_0/{2^j}}\left( {j = 1,2,3} \right)$    in the figure represents the operation $ e^{i{\bf{A}}t_0/{2^j}} \otimes {{\bf{I}}_n}$.}
\label{fig5}
\end{figure}

    We demonstrate a proof-of-principle experiment of the QSVT algorithm shown in Fig. \ref{fig5}. This simple quantum circuit solves a meaningful instance of the problem, that is, to perform the SVT on a $2 \times 3$ dimension matrix ${\bf{A}}_0$.
    For the numerical example, we select different inputs of ${\bf{A}}_0$ all satisfying that the singular values of ${\bf{A}}_0$ are ${\sigma _1} = 2,{\sigma _2} = 1$.
    Matrix ${\bf{A}}_0$ is selected such that the eigenvalues of ${\bf{A}}$ are 4 and 1, which can be exactly encoded with three qubits in register $C$.
    Without loss of generality, let $\tau=1/2$.
    This allows us to optimize the subroutine ${\bf{U}}_{\sigma,\tau}$ (in Fig. \ref{fig1}) without involving register $L$ and ancilla qubits $\left| \tau  \right\rangle$.
    The input state of register $B$ is a normalized quantum state $\left| \psi_{{\bf{A}}_0} \right\rangle $.

    	The initial quantum system is $\left| 0000b \right\rangle$.
	Phase estimation generates the states that encodes the eigenvalues of ${\bf{A}}$ in register $C$, and subsequently, the system is in the superposition :  $\frac{1}{{\sqrt 5 }} (2\left| {0100} \right\rangle \left| {{{\bf{u}}_1}} \right\rangle \left| {{{\bf{v}}_1}} \right\rangle  + \left| {0001} \right\rangle \left| {{{\bf{u}}_2}} \right\rangle \left| {{{\bf{v}}_2}} \right\rangle )$.	
	The mapping of the operator ${\bf{U}}_{\sigma,\tau}$ is:
	$\left| {100} \right\rangle^C  \mapsto \left| {110} \right\rangle^C $
	and
	$\left| {001} \right\rangle^C  \mapsto \left| {100} \right\rangle^C $
	where the outputs $\left| {110} \right\rangle$  and $\left| {100} \right\rangle$ can be interpreted as the encodings ${2^3} \left( {1 - \tau/{\sigma _1}} \right) = 6$ and ${2^3}  \left( {1 - \tau/{\sigma _2}} \right) = 4$ respectively.
	 After the ${\bf{U}}_{\sigma,\tau}$ operator, the system becomes $\frac{1}{{\sqrt 5 }} (2\left| {0110} \right\rangle \left| {{{\bf{u}}_1}} \right\rangle \left| {{{\bf{v}}_1}} \right\rangle  + \left| {0100} \right\rangle \left| {{{\bf{u}}_2}} \right\rangle \left| {{{\bf{v}}_2}} \right\rangle )$.	
	Then we use ${\left| {{2^3} \left( {1 - \tau /{\sigma _k}} \right)} \right\rangle }$  in  register $C$ as the control register to execute a sequence of Pauli $Y$ rotations ${{\bf{R}}_y}$ on the ancilla qubit $\left| 0 \right\rangle$ with $\alpha = \frac{\pi}{2/(1-\tau/\sigma_1)}=2.0944$.
	Take the inverse of all the operations before ${{\bf{R}}_y}$,  measure the ancilla qubit to be $\left| 1\right\rangle$, the system now becomes
	\begin{equation}
        \begin{aligned}
    		\frac{1}{{\sqrt {N_\alpha} }}\left| {1000} \right\rangle \left( {1.9999 \left| {{u_1}} \right\rangle \left| {{v_1}} \right\rangle  + 0.8660 \left| {{u_2}} \right\rangle \left| {{v_2}} \right\rangle } \right),
	\end{aligned}
	\end{equation}
	where $N_\alpha  = 4.7495$.

	The theoretical result is as follows:
        \begin{equation}
        \begin{aligned}
    		\frac{1}{\sqrt{N_2}} \sum\nolimits_{k = 1}^2 {\left({\sigma _k-\tau}\right)\left| 1000\right\rangle \left| {{{\bf{u}}_k}} \right\rangle \left| {{{\bf{v}}_k}} \right\rangle } 
		= \frac{1}{{\sqrt {10} }} \left| {1000} \right\rangle( 3 \left| {{{\bf{u}}_1}} \right\rangle \left| {{{\bf{v}}_1}} \right\rangle  + \left| {{{\bf{u}}_2}} \right\rangle \left| {{{\bf{v}}_2}} \right\rangle ).
	\end{aligned}
	\end{equation}

	We then compute the probability and the fidelity according to the Eqs. (\ref{eq_P}) and (\ref{eq_F}). The probability of measuring the ancilla qubit to be 1 is  $P(\alpha) = 0.9499$, and the fidelity is  $F(\alpha) = 0.9962$.
	
    	Our results may motivate experimentalists to verify this result by implementing the quantum circuit  with capability of addressing 6 or more qubits and execute basic quantum gates on their setups.

\section{Conclusions}
\label{sec5:level1}

Nowadays, the quantum circuit model is the most popular and developed model for universal quantum computation. We further investigated the QSVT algorithm which we proposed in Ref. \cite{Duan17} by providing the possibility to implement the algorithm on a quantum computer via the circuit model. The scalable quantum circuit is presented, in which the key subroutine of the controlled rotation is designed by introducing Newton's method and an adjustable parameter $\alpha$. We simplified the Newton iteration function in terms of an intermediate variable and reduce the number of Newton iterations with the help of a magic initial state. Then we analyzed the space/time complexity of the designed quantum circuit which shows that the number of qubits and gates required in the quantum circuit are $O\left[ {{{\log }_2}\left( {pq} /{\varepsilon} \right)} \right]$ and $O\left[ {ploy\log_2 \left( {1/\varepsilon } \right)} \right]$ respectively. Moreover, we provided two methods to compute the value of $\alpha$ to ensure high probability and high fidelity readout and conducted numerical experiments. The numerical results show that under different inputs, our method can output
 high probability and high fidelity in one iteration of the quantum circuit. Furthermore, we present a small-scale circuit as an example to verify the algorithm. We hope that our research motivates experimentalists to conduct new investigations in quantum computation.

\appendix

\section {Initial state of the Newton's method}
\label{A0}

        For the single-precision floating-point format, the magic number is chosen as 
a hexadecimal constant  $0{\rm{x}}5f375a86$, and for the double-precision floating-point format, the magic number can be chosen as a hexadecimal constant $0{\rm{x}}5fe6ec85e7de30da$.
 As mentioned in Sec. \ref{sec:U}, both the magic numbers follow the IEEE 754 floating-point format. The test in Ref. \cite{INVSQRT} shows that the relative error can be around 0.00175 after 1 Newton iteration, and reduce to around $4.65437e-004$ after 2 Newton iterations. Therefore, our quantum circuit only needs one or two iterations to get the high precision approximation of $z_k$ with the help of the magic initial state.

\section {Taylor's series method}
\label{A1}

Using Taylor's series method to solve Eq. (\ref{eq_G'}), we have
	
	\begin{equation}
	\begin{aligned}
        &G'\left( \alpha  \right) = 0  \\
        & \Rightarrow \sum\nolimits_{k = 1}^r {\sigma _k^2\left( {y_k^2 - \frac{{y_k^4{\alpha ^2}}}{{2!}} + \frac{{y_k^6{\alpha ^4}}}{{4!}} -   \cdots  + {{\left( { - 1} \right)}^n}\frac{{y_k^{2n + 2}{\alpha ^{2n}}}}{{\left( {2n} \right)!}} + o\left( {y_k^{2n + 3}{\alpha ^{2n + 1}}} \right)} \right)}  = 0.
        \end{aligned}
        \end{equation}

If the 2-order approximation is selected, then

	\begin{equation}
	\begin{aligned}
            	&\sum\nolimits_{k = 1}^r \sigma _k^2\left( {y_k^2 - \frac{{y_k^4{\alpha ^2}}}{2} + o\left( {y_k^5{\alpha ^3}} \right)} \right) = 0 \\
            	&\Rightarrow \sum\nolimits_{k = 1}^r {\sigma _k^2\left( {y_k^2 - \frac{{y_k^4{\alpha ^2}}}{2}} \right)}  \approx 0
            	\Rightarrow \alpha  \approx \sqrt {\frac{{2\sum\nolimits_{k = 1}^r {\sigma _k^2 y_k^2} }}{{\sum\nolimits_{k = 1}^r {\sigma _k^2 y_k^4} }}}.
	\end{aligned}
        \end{equation}

If the 4-order approximation is selected, then
	\begin{equation}
	\begin{aligned}
                   & \sum\nolimits_{k = 1}^r {\sigma _k^2\left( {y_k^2 - \frac{{y_k^4{\alpha ^2}}}{2} + \frac{{y_k^6{\alpha ^4}}}{{4!}} + o\left( {y_k^7{\alpha ^5}} \right)} \right)}  = 0  \\
                   & \Rightarrow \sum\nolimits_{k = 1}^r {\sigma _k^2\left( {y_k^2 - \frac{{y_k^4{\alpha ^2}}}{2} + \frac{{y_k^6{\alpha ^4}}}{{4!}}} \right)}  \approx 0
                   \Rightarrow \alpha  \approx \sqrt {\frac{{b - \sqrt {{b^2} - 4ac} }}{{2a}}},
	\end{aligned}
    \end{equation}
where $a = \frac{1}{{24}}\sum\nolimits_{k = 1}^r {\sigma _k^2  y_k^6} ,b = \frac{1}{2}\sum\nolimits_{k = 1}^r {\sigma _k^2  y_k^4} ,c = \sum\nolimits_{k = 1}^r {\sigma _k^2  y_k^2} $.

\section{Numerical simulations}
\label{A2}
    \emph{Initialization}. Recall that the applications of SVT are mainly based on videos or pictures. Thus, we select 120 different inputs of matrix
     ${\bf{A}}_0^{\left(i\right)} (i = 1,2,...,120)$ ,
    which are derived from random pictures.
    These matrices have been pre-processed in terms of the condition number and the $\ell_2$ normalization. The efficient condition number $\kappa$ of the matrix ${{\bf{A}}^{\left(i\right)}}={\bf{A}}_0^{\left(i\right)}{\bf{A}}_0^{{\left(i\right)}\dag}$ are set to $10^6$, therefore, only the singular values in the range of  [1/1000,1] are taken into account.
    The $\ell_2$ normalization ${\left\| {\bf{A}}_0^{\left(i\right)} \right\|_2} $ are set to 1, therefore, the largest singular value ${\sigma_1^{\left(i\right)}}$  of each matrix equals to 1. (Alternatively, Frobenius norm can also be used to normalize these matrices.)
    The dimensionality of these matrices ranges from $200 \times 300$ to $2621 \times 3995$, and the rank of these matrices ranges from 55 to 668.
 The singular value distributions of the 120 matrices ${\bf{A}}_0^{\left(i\right)}$ are shown in Fig. \ref{fig:svd}.
    Typically, the largest singular value dominates the most part of all singular values, and only few principle singular values of the matrix ${\bf{A}}_0^{\left(i\right)}$ are large, while the rest amount of the singular values are very small.
	In the following experiments, all the singular values of the 120 matrices are computed, and the probability and fidelity are then derived by Eq.(\ref{eq_P}) and Eq.(\ref{eq_F}), respectively.
 
{\color{red} 
\begin{figure}[ht]
\centering
\includegraphics[width=0.6\linewidth]{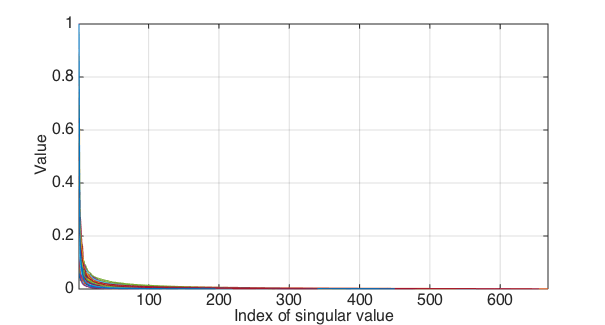}
\caption{Singular value distributions of the 120 matrices.}
\label{fig:svd}
\end{figure}
 }

\begin{figure}[htbp]
    \centering
    \subfigure{
    \begin{minipage}{8cm}
    \centering
    \includegraphics[width=1\linewidth]{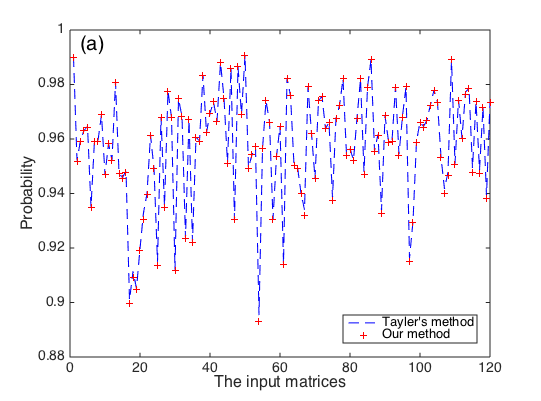}
    \end{minipage}}
    \subfigure{
    \begin{minipage}{8cm}\centering
    \includegraphics[width=1\linewidth]{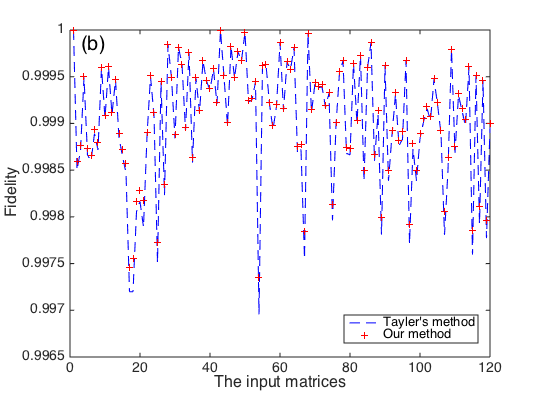}
    \end{minipage}}
    \caption{Probability and fidelity based on the solutions obtained by Taylor's method and our method. (a) Probability. (b) Fidelity.} 
    \label{fig:pft}
    \end{figure}

    {\bf{ Experiment(1).}} Fig. \ref{fig:pft} shows that the $x$ axis represents the 120 inputs. Fig. \ref{fig:pft}(a) and (b) show the probability and the fidelity in terms of different $\alpha$, which are obtained by Taylor's method (in blue dashed line) and our `intuitive' method (in red '+'), respectively. 	
    The numerical results show that our `intuitive' method works well as the Taylor's method.
    The probability is almost the same in terms of $\tilde \alpha$ obtained by our method and Taylor's method. Moreover, fidelity performs slightly better when our method is used instead of the Taylor's method.
    Both methods can help output the high probability and high fidelity in the context of the 120 different random inputs.

    {\bf {Experiment(2).}} We now analysis which properities of the input matrix ${\bf{A}}_0^{\left(i\right)} $ and hyperparameter $\tau$ impact the probability and fidelity.

\begin{figure}[htbp]
    \centering
    \subfigure{
    \begin{minipage}{16cm}
    \centering
    \includegraphics[width=1\linewidth]{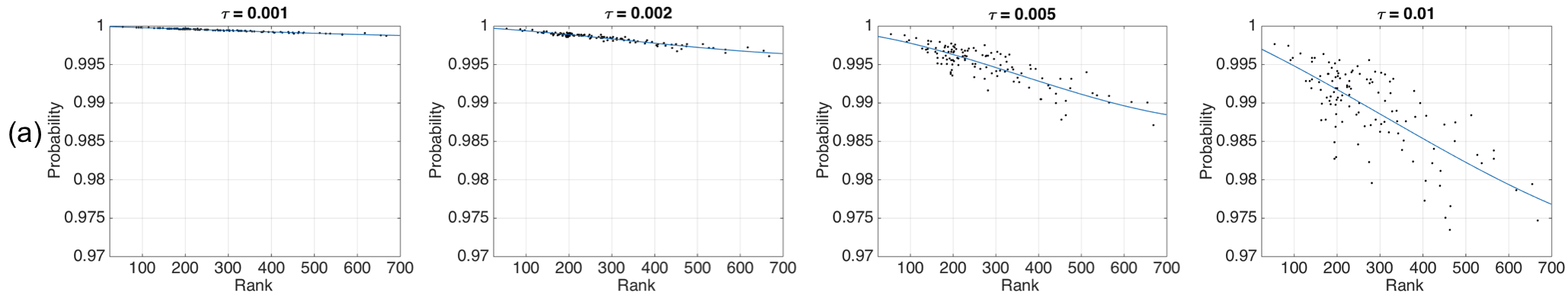}
    \end{minipage}}
    \subfigure{
    \begin{minipage}{16cm}\centering
    \includegraphics[width=1\linewidth]{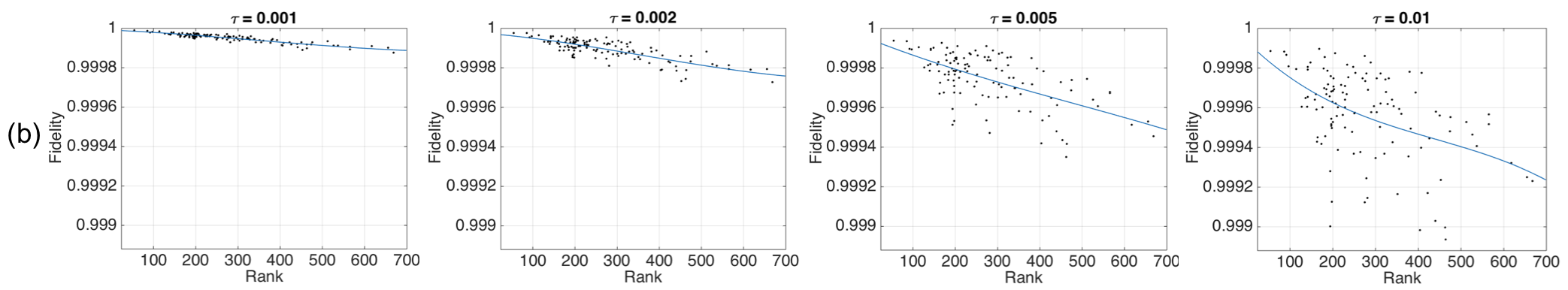}
    \end{minipage}}
    \caption{Probability and fidelity with the rank of input matrices. (a) Probability. (b) Fidelity.}
    \label{fig:pfr}
    \end{figure}

    Specifically, we select the hyperparameter $\tau$ from \{0.001, 0.002, 0.005, 0.007, 0.009, 0.01, 0.02, 0.05, 0.07, 0.09, 0.1, 0.2, 0.3, 0.4, 0.5\}. Fig. \ref{fig:pfr} shows how the probability and the fidelity changes with the rank of the input matrices. Here we pick up the pictures with significant features to show up the trends. The black dots represent the probability or fidelity in Fig. \ref{fig:pfr} (a)  and (b) respectively, and the blue solid lines represents the 3-order polynomial fit function generated via cftool in Matlab.

    Fig. \ref{fig:pfr} shows that the rank $r$ of the input matrices would impact the probability and fidelity of the readout, but the effect is not significant. When $\tau$ is small, the probability/fidelity decreases while the rank of input matrices increases. As the $\tau$  become large, the relationship of probability/fidelity and the rank become unapparent. We can also see that probability is more susceptible to rank than fidelity. The results are reasonable. According to the Eqs. (\ref{eq_P}) and (\ref{eq_F}) which derive the probability and fidelity, the rank of the input matrix would affect the total number of summed items in the numerator and denominator at the same time. But large amount of the items are very small since many singular values of the input matrix are small. Therefore, the probability and fidelity may not be significantly affected by the rank of the matrix.

    \begin{figure}[htbp]
    \centering
    \subfigure{
    \begin{minipage}{16cm}
    \centering
    \includegraphics[width=0.85\linewidth]{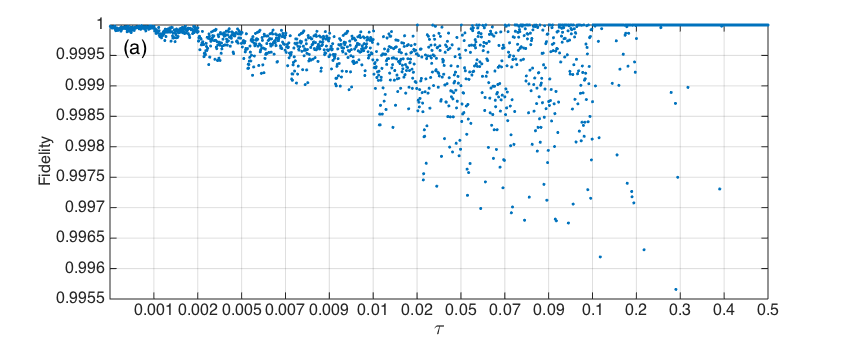}
    \end{minipage}}
    \subfigure{
    \begin{minipage}{16cm}\centering
    \includegraphics[width=0.85\linewidth]{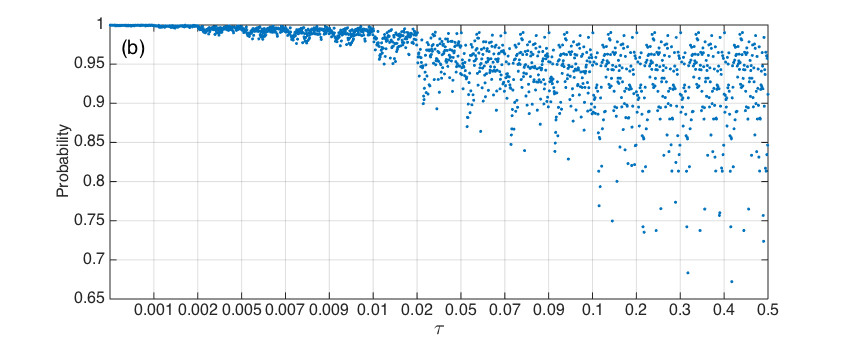}
    \end{minipage}}
    \subfigure{
    \begin{minipage}{16cm}\centering
    \includegraphics[width=0.85\linewidth]{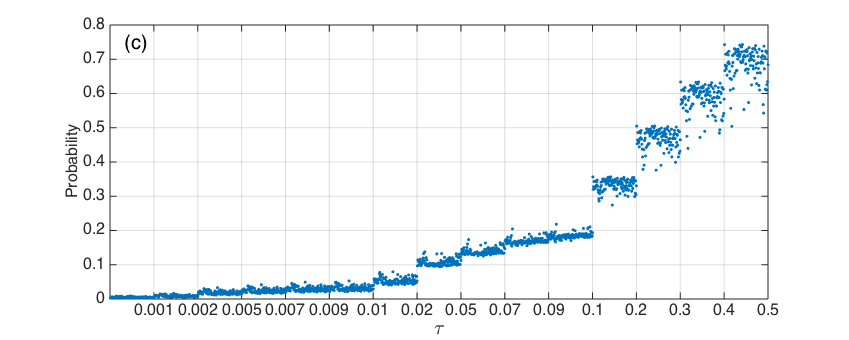}
    \end{minipage}}
    \caption{(a) Fidelity with different input matrices and different $\tau$. (b) Probability with different input matrices and different $\tau$. (c) Difference of the probabilities in terms of our method and the normal method with different input matrices and different $\tau$.}
    \label{fig:fpp}
    \end{figure}

    Fig. \ref{fig:fpp} shows how the probability and the fidelity change with the hyperparameter $\tau$. Each label $\tau$ in the $x$ axis consists of 120 inputs (i.e. the dimensional of $x$ axis amounts to $15 \times 120$). The blue dots in Fig. \ref{fig:fpp} (a) and (b) represent the fidelity and the probability derived by Eq.(\ref{eq_F}) and  Eq.(\ref{eq_P}) respectively. The blue dots in Fig. \ref{fig:fpp} (c) represent the difference between the probabilities derived by our method and the method in Ref. \cite{CD16}, i.e. $\Delta P = P(\alpha) - P_1$.

    As shown in Fig. \ref{fig:fpp}(a), the fidelity is higher than $99.95\%$ when $\tau \le 0.002$. Then the fidelity declines and falls to a lowest point, at which point it begins to increase, reaching its highest point at 1 (when ${\sigma _2} \le \tau  < {\sigma _1}$, 
the fidelity is $F\left( \alpha  \right) = {{\sigma _1^2{y_1}} \mathord{\left/
 {\vphantom {{\sigma _1^2{y_1}} {\sqrt {\sigma _1^2y_1^2 \times \sigma _1^2} }}} \right.
 \kern-\nulldelimiterspace} {\sqrt {\sigma _1^2y_1^2 \times \sigma _1^2} }} = 1$ according to Eq. (\ref{eq_F})).  
    The lowest point among the 120 different matrices is just over $99.55\%$, which shows that our method can derive a high fidelity even at the worst case.

    As shown in Fig. \ref{fig:fpp}(b), the probability decreases steadily while $\tau$ rises. This trend can be proved by Theorem 1 below. When $\tau \le 0.02$, the probability can be reached to the point over $95\%$. And the lowest probability is over $65\%$ among the 120 different matrices when $\tau$ is large.
    Comparing to Fig. \ref{fig:fpp}(c),  the difference between the probabilities derived by our method and the method in Ref. \cite{CD16} increases steadily while $\tau$ rises.
    When $\tau \le 0.02$, the difference is less than $10\%$.  At the point $\tau =0.1$, the difference begins to increase rapidly. This infers that when $\tau$ is large, our methods can derive much higher probability than the normal method.

    The probability of obtaining the final result would affect the iteration of the quantum circuit. Theorem 1 gives a lower bound of the probability.

    \textbf{Theorem 1}: \emph{For a given input matrix ${\bf{A}}_0$, the probability of obtaining the final result is a non-increasing function of the hyperparameter $\tau$. And the lower bound of the probability is }
	 	\begin{equation}
            P_{min}=\frac{{{\sigma _1}}}{{\sum\nolimits_{k = 1}^r {\sigma _k^2} }}.
            \label{Pmin}
		\end{equation}
    Apparently, the iteration of the quantum circuit is $O\left( {{{\sum\nolimits_{k = 1}^r {\sigma _k^2} }}/{{{\sigma _1}}}} \right)$.

    \textbf{Proof:} According to Eq. (\ref{eq_P}) and the value of $\tilde \alpha = \frac{\pi}{2\left(1-\tau/\sigma_1 \right)}$, the probability becomes:
    \begin{equation}
        P = \frac{{\sum\limits_{k = 1}^r {\sigma _k^2{{\sin }^2}\left( {{y_k}\alpha } \right)} }}{{\sum\limits_{k = 1}^r {\sigma _k^2} }} = \frac{{\sum\limits_{k = 1}^r {\sigma _k^2{{\sin }^2}\left[ {\frac{{\pi {\sigma _1}}}{{2{\sigma _k}}} \cdot \frac{{{{\left( {{\sigma _k} - \tau } \right)}_ + }}}{{{\sigma _1} - \tau }}} \right]} }}{{\sum\limits_{k = 1}^r {\sigma _k^2} }}  \buildrel \Delta \over = P\left( \tau  \right).
    \end{equation}
    Obviously, as ${\sigma _k} \le {\sigma _1}$ and $\tau > 0$, the function $f\left( \tau  \right) = {\frac{{\pi {\sigma _1}}}{{2{\sigma _k}}} \cdot \frac{{{{\left( {{\sigma _k} - \tau } \right)}_ + }}}{{{\sigma _1} - \tau }}}$ is a non-increasing function of the hyperparameter $\tau$.
    And as $y_1\alpha \in [0,\pi/2]$ which has been assumed in Sec. \ref{sec:PF}, the function $g\left( \tau  \right) = {\sin ^2}\left( {f\left( \tau  \right)} \right)$ is an increasing function of $f\left( \tau  \right)$. Therefore, $P\left( \tau  \right)$ is a non-increasing function of $\tau$. Specifically, when $\tau$ satisfies ${\sigma _2} < \tau  < {\sigma _1}$, $ P\left( \tau  \right)$ get the lower bound shown as in Eq. (\ref{Pmin}).

\begin{acknowledgments}
We would like to thank Patrick Rebentrost, Maria Schuld, Shengyu Zhang and Chaohua Yu for helpful discussions. This work is supported by NSFC (Grant Nos. 61571226, 61572053, 61701229, 61702367), Natural Science Foundation of Jiangsu Province, China (Grant No. BK20170802), the Beijing Natural Science Foundation (Grant No. 4162005), Jiangsu postdoctoral science foundation, China Postdoctoral Science Foundation fund, the Research Project of Tianjin Municipal Commission of Education(Grant No. 2017KJ033).\end{acknowledgments}

\bibliography{QSVT}

\end{document}